\input harvmac.tex
%\draftmode
\noblackbox
\def\ap{\alpha'}
\Title{\vbox{\baselineskip12pt
\hbox{\tt CALT-68-2298, CITUSC/00-055}
\hbox{\tt hep-th/0009181}}}%
{\vbox{\centerline{Non-Relativistic Closed String Theory}}}
\bigskip
\medskip
\centerline{Jaume Gomis and Hirosi Ooguri}
\bigskip
\centerline{California Institute of Technology 452-48, Pasadena, CA 91125}
\centerline{{\it and}}
\centerline{Caltech-USC Center for Theoretical Physics, Los Angeles, CA}
\medskip
\centerline{\tt gomis,\ ooguri@theory.caltech.edu}
\medskip
\medskip
\bigskip
\bigskip
\bigskip
\centerline{{\bf Abstract}}
\bigskip

We construct a Galilean invariant non-gravitational closed string
theory whose 
excitations satisfy a non-relativistic dispersion relation. This
theory can be obtained by taking a consistent low energy limit of any of the
conventional string theories, including the heterotic string. We give
a finite first order worldsheet Hamiltonian for this theory and show
that this string theory has a sensible perturbative expansion,
interesting high   energy behavior of scattering amplitudes and a
Hagedorn transition of the thermal ensemble. The strong coupling duals
of the Galilean superstring theories are considered and are shown to be
described
by an eleven-dimensional Galilean invariant theory 
of light membrane fluctuations. A  new class of 
Galilean invariant non-gravitational theories of light-brane excitations are
obtained. We exhibit dual formulations of the strong coupling limits
of these Galilean invariant theories and show that they exhibit many
of the conventional dualities of M theory in a non-relativistic setting.

\Date{September 2000}

\newsec{Introduction}

One of the legacies of the second superstring revolution is the
realization that the different superstring theories describe very
special corners of the space of vacua of a single hypothetical structure
dubbed M Theory. Another important lesson that has emerged is
that there are regions of the space of vacua describable by a theory
without gravity. Two beautiful examples of such theories are 
Matrix Theory \ref\matric{T. Banks, W. Fischler, S.H. Shenker, and 
L. Susskind, ``M Theory as a Matrix Model: a Conjecture,''
Phys. Rev. D55 (1997) 5112, {\tt hep-th/9610043}.} 
and Maldacena's conjecture \ref\malda{J. Maldacena, ``The Large $N$
Limit of Superconformal Field Theories and Supergravity,''
Adv. Theor. Math. Phys. 2 (1998) 231, {\tt hep-th/9711200}.}. 
The realization that
there are consistent limits of M Theory without gravity has  led to a
geometrical 
understanding of some field theory dualities and to new, hitherto
unknown, field theories in higher dimensions.

The non-gravitational limits studied thus far involve considering
certain low energy limits of M Theory in the presence of
branes. Typically, these limits lead to a theory where the  appropriate
effective description is given in terms of the  massless degrees of
freedom propagating on the 
branes. Such low energy limits lead, for example, to gauge theories in
various dimensions. In such examples, the massive open string states
on the branes and the entire closed string spectrum decouple
from the low energy physics and the truncation to the theory of the massless
fluctuations is consistent. These low energy theories are described by
field theories.

Recently, very interesting generalizations have been found in which
 closed strings decouple but the massive open string
excitations on the branes need to be taken into account for physical
processes
\lref\sst{N. Seiberg, L. Susskind and N. Toumbas,
``Strings in Background Electric Field, Space/Time Noncommutativity
and a New Noncritical String Theory,'' JHEP 0006 (2000) 021, 
{\tt hep-th/0005040}.} \lref\gmms{R. Gopakumar, J. Maldacena,
S. Minwalla, and A. Strominger, ``S Duality and Noncommutative
Gauge Theory,'' JHEP 0006 (2000) 036, {\tt hep-th/0005048}.}
\refs{\sst,\gmms}. These
theories appear in low energy limits of branes in near critical electric
field backgrounds and are not conventional field theories due to
the presence of a tower of  massive excitations. Since massive states cannot
be neglected, the field theory truncation is not unitary
\lref\gm{J. Gomis and T. Mehen, 
``Spacetime Noncommutative Field Theories and Unitarity,''
{\tt hep-th/0005129}.}\lref\revised{N. Seiberg, L. Susskind,
and N. Toumbas, ``Spacetime Noncommutativity and Causality,''
JHEP 0006 (2000) 044, {\tt hep-th/0005015}.}
\lref\agm{O. Aharony, J. Gomis and T. Mehen, ``On Theories with
Light-Like Noncommutativity,'' {\tt hep-th/0006236}.}
\refs{\gm,\revised,\agm}. 
These non-gravitational theories describe   
all the fluctuations  on the branes. For example, 
 one can obtain a consistent open
string theory without any closed string states.
Such theories arise from studying D-branes in a
background electric field (NCOS) \refs{\sst,\gmms}, 
M5-branes in a three-form 
background (OM)\lref\gmss{R. Gopakumar, S. Minwalla, N. Seiberg and
A. Strominger, ``OM Theory in Diverse Dimensions,'' JHEP 0008 (2000)
 008, {\tt hep-th/0006062}.} \gmss\
 and Neveu-Schwarz five-branes in various constant 
Ramond-Ramond $p$-form backgrounds (OD) 
\lref\bbss{
E.Bergshoeff, D.S. Berman, J.P. van der Schaar, and
P. Sundell, ``Critical Fields on the M5-brane and Noncommutative
Open Strings,'' {\tt hep-th/0006112}.}
\lref\har{T. Harmark,
``Open Branes in
Space-Time Non-Commutative Little String Theory'', {\tt hep-th/0007147}.}
\refs{\gmss,\bbss,\har}. 
\lref\km{I.R. Klebanov and J. Maldacena, ``$(1+1)$-Dimensional
NCOS and Its $U(N)$ Gauge Theory Dual,''
{\tt hep-th/0006085}.}

In this paper we find that there are corners of the moduli space of vacua of
M Theory without branes that are described by non-gravitational theories whose
excitations live in space-time. These massive excitations satisfy
a non-relativistic dispersion relation and the theory that describes
their dynamics is unitary and has a sensible perturbative description
(whenever one  is available). Since background branes
are not required to define these non-relativistic theories, 
they can be 
obtained by taking certain low energy limits of all five superstring
theories, including the heterotic string. We will call these theories
non-relativistic closed string theories (NRCS).

The simplest limit leading to NRCS is obtained by considering string theory
in the presence of a near critical NS-NS two-form field without any
D-brane\foot{When the NS-NS two-form exceeds the critical value,
the space-time energy of a closed string becomes unbounded below
and can become indefinitely negative as we increase the winding number.}.
In the context of $(1+1)$-dimensional NCOS, Klebanov and Maldacena
\km\ observed
that when the spatial direction is compactified on a circle, that there
are finite 
energy winding closed string states that do not decouple from the
open strings. An example of NRCS can be obtained by considering
precisely the NCOS limit without any D-brane. 
Naively, one might think that, in the absence of D-branes, that a constant
NS-NS two-form can be 
gauged away and that one ends up getting a conventional relativistic closed
string theory. 
%However, if the B field is along a compact direction,
%the B field shifts the Hamiltonian of the theory in the sector where the
%strings have non-zero winding number. 
%By appropriately tunning the B
%field along the circle and taking a low energy limit one can get
%non-relativistic 
%closed strings. 
%This would indeed be the case before the NCOS limit.
%Once the limit is taken and the spectrum is truncated, however,
%the theory is forever altered. 
This is obviously true in non-compact space. However, in the presence
of a circle, the background NS-NS field modifies the spectrum, which
remains relativistic. Once the NCOS limit is taken, there is a
truncation of the low energy spectrum and one obtains a new theory with a
Galilean invariant Hamiltonian.
Perhaps surprisingly, the closed
string theory in the NCOS limit without any D-brane has
a well-behaved perturbative expansion, described by the Lagrangian in
section $3$. It is also interesting to study  the worldsheet theory we
propose when the worldsheet has a boundary. Then, our formalism
reproduces the relativistic open string spectrum of NCOS and its
interactions.

In section $3$, we give a worldsheet Lagrangian
for NRCS, which has Galilean 
invariance and from which we derive the non-relativistic
spectrum of closed strings and
their interactions. The Lagrangian we propose can be derived from the
conventional Polyakov path integral quantization of the relativistic
string by rewriting it in variables that are conducive to taking the
low energy limit that defines NRCS (see section $3$ for details of the
limit). We explicitly solve the Virasoro
constraints, thus yielding the spectrum, show that the
theory is unitary and that it has a sensible perturbative
expansion. The string spectrum, 
being non-relativistic, does not contain a massless
graviton and it is thus non-gravitational in nature. However, there is
an instantaneous Newtonian potential 
between the massive strings. This string theory exhibits interesting
properties such as an unusual high energy behavior of scattering
amplitudes and a Hagedorn transition of the thermal ensemble.

NRCS depends on two-parameters, the effective string scale
$\alpha_{eff}'$ and the effective string coupling constant $g$. One may ask
what is the strong coupling dual of these theories. For the
superstrings, this can be reliably answered. We find that the strong
coupling limits of supersymmetric NRCS are given by a Galilean
invariant  eleven
dimensional theory of light membranes which we call GM (Galilean
membrane theory). This eleven dimensional theory
has a unique dimensionfull parameter $l_{eff}$  which is the effective
Planck length. The relation between the NRCS superstrings and GM is
reminiscent to the relation between the conventional superstrings and
M theory. For example, Type IIA NRCS with coupling $g$ and string
scale $\alpha_{eff}'$ is equivalent to GM on a circle of radius $R$ such
that $R=g\sqrt{\alpha_{eff}'}$ and $l_{eff}=g^{1/3}\sqrt{\alpha_{eff}'}$. The
conventional dualities  and relations with M theory still hold, such
that, for example, Type IIB NRCS has an $SL(2,Z)$ symmetry. We discuss
these relations in section $7$. It is interesting that duality
symmetries in string theories do not rely on relativistic invariance
nor the presence of gravity.

There are many interesting generalizations that can be made that lead
to non-relativistic, non-gravitational theories. The construction of
such theories is quite general. The basic idea is to study the low
energy limit of M Theory vacua in the presence of any of the many
possible gauge fields available. Then, one can take a low energy, near
critical limit
such that all states of M Theory become infinitely massive, and thus
decouple, except for those states that couple to the constant near
critical background gauge field. Tuning to the critical value, defined such
that that energy coming from the background field precisely cancels
the rest energy of the states in question, ensures that even though we
are taking a low energy limit, that there are states that survive
and
satisfy a non-relativistic dispersion relation. For example, if we
tune the background NS-NS B field to its critical value, then one
obtains finite energy non-relativistic fluctuations of strings winding
around the circle. Clearly, such NRCS can be defined in Type II, Type I and
Heterotic theories. Moreover, if one considers, for example, a near
critical R-R gauge field $C_{p+1}$ and takes a low energy limit, 
then there are light Dp-branes\foot{In order for the constant
background $C_{p+1}$ field to affect the energy of a Dp-brane, the
brane has to be wrapped on a $p$-cycle, otherwise the gauge field can
be gauged away without changing the energetics.}
 which are non-relativistic that
decouple from all the rest of the modes and lead to decoupled Galilean
invariant theories which we will call GDp (Galilean Dp-brane
theories). The 
myriad of gauge fields that exist in M Theory
vacua can be used to define new non-gravitational Galilean invariant
theories. We study such theories in section $7$.
The dualities of the underlying relativistic M Theory, lead to
interesting webs of dualities for these non-relativistic theories.
These non-relativistic theories may be  a promising ground in
which to address some of the important questions of M Theory without
the complication of gravity.

The rest of the paper is organized as follows:

In section $2$ a very
general low energy limit is presented which yields a finite non-relativistic
dispersion  
relation from the spectrum of a  charged relativistic brane. The
limit, when 
applied to the fundamental closed string, yields  the spectrum of
NRCS. Generalizations to other relativistic objects in M Theory are
briefly described.

In section $3$ we find the worldsheet theory of
NRCS. We quantize the Galilean invariant, first-order Hamiltonian and
find under what conditions there is a physical closed string
spectrum. We then reproduce the NRCS spectrum in section $2$ within our
Hamiltonian formalism. We compute the BRST cohomology of the string
and show that there are no ghosts in the spectrum.
 The possibility of adding a boundary to the
worldsheet is considered. The formalism of section $3$ results in
the spectrum and worldsheet correlation functions of NCOS.
Using  this formalism, it is straightforward to prove 
the decoupling of the massless
open string states on worldsheets with any number of handles and holes
when the longitudinal direction is non-compact. This
extends the result of \km\ to all orders in the perturbative
expansion.

Section $4$
is devoted to performing tree level computations in NRCS. We show that
scattering amplitudes have the correct pole structure required by
unitarity and have a peculiar behavior of high energy fixed
angle scattering amplitudes in NRCS. Despite the absence of gravity in
this theory, we exhibit a Newtonian potential among the
non-relativistic strings. 

In section $5$ we compute loop amplitudes
and show that NRCS is a sensible theory in perturbation theory. We
evaluate the Helmholtz free energy at one loop
and reproduce from it the spectrum of NRCS
found in section $2$. We find that NRCS behaves
 similarly
to the long string near the boundary of $AdS_3$
\ref\mos{J. Maldacena, H. Ooguri, and J. Son,
``Strings in $AdS_3$ and the $SL(2,R)$ WZW Model. Part 2:
Euclidean Black Hole,'' {\tt hep-th/005183}.}.
We exhibit the existence of a Hagedorn
temperature in NRCS and sketch higher loop computations.
We also study in some detail $N$-point loop
amplitudes and show that the amplitudes are finite.

In section $6$ we elucidate the relation between NRCS and the discrete
light-cone quantization (DLCQ) of closed string theory. NRCS is
related by T-duality 
to the discrete light-cone quantization (DLCQ) of closed string theory. 
Therefore, the formalism developed in this paper provides a useful
description of DLCQ string theories as well. 

In section $7$ we study Galilean invariant theories of light branes
and some of their dualities. In eleven dimensions we study the
Galilean invariant theory of membrane fluctuations (GM) and five-brane
fluctuations (GF). In ten dimensions we discuss the theory of
non-relativistic light Dp-branes (GDp) and light Neveu-Schwarz
five-branes (GNS). These theories lie in the same moduli space and
exhibit the same dualities that the underlying relativistic M theory
possesses. In particular we show that the strong coupling limits of
some NRCS have an eleven dimensional description in terms of light
brane excitations.

\newsec{Non-Relativistic Limit}

In this section we show that by taking a low energy limit of the
theory of a relativistic $p$-brane and by tuning the $p+1$ gauge field
that couples to it, that  one can obtain an exact non-relativistic
dispersion relation. The idea is to study the low energy spectrum in
a scaling limit in which the background gauge field cancels the rest
energy of the brane  and such that the non-relativistic approximation
becomes exact. 
In this limit, all the states of the theory decouple, except the light
p-brane excitations.
We present the truncation to a non-relativistic theory in a
very simple toy model which captures the essence of the limit which
defines NRCS and the other generalizations we describe in this paper.

For simplicity, consider a relativistic charged point particle of mass $m$ and
charge $e$ coupled to a  gauge field $A_\mu$ propagating in a geometry
with metric components $g_{00}=-1$ and $g_{ij}$ arbitrary with $i,j\neq 0$.
The Lagrangian which describes its motion is given by
\eqn\pplag{ L = -m \sqrt{-\dot x^2} + e A_\mu \dot x^\mu .}
Worldline reparametrization invariance implies
 Einstein's dispersion relation
\eqn\einstein{ p_0 = 
\sqrt{m^2 + g^{ij}(p_i-eA_i)(p_j-eA_j) } + e A_0.}
Consider the following low energy limit 
\eqn\lowener{
g_{ij}={m_{eff}\over m}\delta_{ij} \quad eA_0=-m+ e\hat{A}_0}
as $m\rightarrow \infty$. In this limit, Einstein's relation
\einstein\ 
reduces to the following non-relativistic dispersion law
\eqn\nonr{
p_0={1\over 2m_{eff}}(p_i-eA_i)^2+e\hat{A}_0.}
Although a constant gauge field can be locally gauged away and does
not affect the equations of motion, it changes
the energy spectrum in the sector of the theory carrying electric
charge. In fact, 
the shift in the energy due to the gauge field precisely cancels the
rest mass of the particle and ensures that the energy remains finite
in the limit
\lowener . Turning on a background field and tuning it to the critical
value is an efficient way of rearranging the spectrum of the theory
such that only states charged under the gauge field have finite
energy, the neutral states acquire infinite proper energy.

The charged point particle model can also be used to show that the
there are finite energy, non-relativistic winding closed string states
in the NCOS limit whenever the near critical NS-NS $B_{01}$ field is along a
compact spatial direction. The mass of a closed string winding
$w$-times around a circle of radius  
$R$ is  
\eqn\mass{
m^2= \left({wR \over \alpha'}\right)^2 + {2(N + \bar{N})\over \alpha'},}
where $N$ and $\bar{N}$ are the amounts of stringy
excitations in the left and the right mover oscillators of the
string. Moreover, the winding string states are charged under the
$U(1)$ gauge field obtained by reducing the NS-NS B field along the
circle. The charge is given by
\eqn\chargerelation{
  e A_0 =  -2 \pi R w B_{0 1}.}
We now take the NCOS limit \refs{\sst,\gmms}\ in the point particle
analogy  \nonr 
\eqn\analoglimit{
g_{ij}  = {\alpha'\over \alpha_{eff}'}\delta_{ij} ,~~ e A_0 = 
-{wR\over\alpha'}  \left( 1 - { \alpha'\over 2\alpha_{eff}'}\right).}
Taking $\alpha' \rightarrow 0$ results in the following
non-relativistic spectrum
\eqn\windingenergyformula{
p_0 = {wR \over 2 \alpha_{eff}'}+{ \alpha_{eff}' k^2\over 2wR}+{N +
\bar{N}\over wR}.}
Thus the NCOS limit can be thought of an example of the non-relativistic
limit \lowener . Note that demanding positive energy states selects
strings winding only in a particular direction.
Indeed, the closed string spectrum
\windingenergyformula\ coincides with the one found by Maldacena and
Klebanov in \km . In the next section we give a Galilean invariant,
finite first-order Hamiltonian that describes these closed strings and verify
that there is a consistent perturbative expansion.

The non-relativistic limit we found in \lowener\ can also be
generalized to any state of M Theory which is charged under a gauge
field. If one considers winding fundamental closed strings in the near
critical
NS-NS $B_{0i}$ background of any of the
known string 
theories, one obtains NRCS theories. But we could have considered any
of the branes of M Theory. One could, for example, consider a wrapped membrane
of M Theory on any two-cycle (say on a two-torus or a two-cycle
of a Calabi-Yau) in a critical three-form background. Then, the
membrane is charged under a gauge field $A_0$ obtained by reducing the
three-form on the two-cycle. In the limit \lowener\ one obtains
a non-relativistic theory without gravity. Likewise, for any other
brane. Just like in NCOS, positivity of the energy selects only those
states which are wrapped in a particular direction, states of opposite
orientation are unphysical. In section $7$ we will find low energy
limits leading to Galilean theories of branes and study their strong
coupling duals.

\newsec{Lagrangian and Quantization}

In this section we construct the worldsheet theory of NRCS
and analyze its spectrum and interactions. We consider a certain low
energy limit of string theory
in a near critical NS-NS B-field. The bosonic worldsheet action which describes
this background is given by\foot{Here the worldsheet and target space
metric are taken to be of Lorentzian signature.}
\eqn\original{  
S_0 = -{1\over 4\pi
\ap}\int d^2\sigma \Big( g_{MN}\partial_a X^M\partial^a X^N-2\pi \ap
B_{MN}\;\epsilon^{ab}\partial_a X^M\partial_b X^N\Big),} 
 where $M,N=0,\cdots,9$ and $a,b=0,1$. NRCS is obtained by choosing
the B field with a time-like and a space-like component.
 Without loss of
 generality we consider a $B_{01}\equiv B$ background.
 NRCS is obtained by taking the following zero slope,
near critical  field limit\foot{This is precisely the NCOS limit of
\refs{\sst,\gmms} but without any D-brane.}
\eqn\critical{
2\pi \alpha'
B_{01}=1-{\ap\over 2\alpha_{eff}'}, \qquad g_{\mu\nu}=\eta_{\mu\nu},\qquad
g_{ij}={\ap\over \alpha_{eff}'}\delta_{ij}, \qquad g_s=g\sqrt{\alpha_{eff}'\over \ap}}
as $\ap\rightarrow 0$ where $\mu,\nu =0,1$ and $i,j=2,\cdots,9$, 
and $\alpha_{eff}'$ is the finite 
effective string scale of NRCS and $g$ its  effective coupling
constant. 

Using 
\eqn\gammacoord{
\gamma = X^0 + X^1~,~~~\bar{\gamma} = -X^0 + X^1}
for
the target space coordinates, 
\eqn\zcoordinates{z = e^{i(\sigma^0 + \sigma^1)}~,~~~
\bar{z} = e^{i(\sigma^0 - \sigma^1)}} 
for the worldsheet coordinates, and
the background given in \critical , the action 
\original\ can be written for finite $\ap$ as 
\eqn\originalnew{
S_0 = -{1\over 4\pi
\ap}\int d^2z\Big(\partial \gamma
\bar{\partial}\bar{\gamma}+\partial\bar{\gamma}\bar{\partial} \gamma
- 2\pi \ap B (\partial \gamma
\bar{\partial}\bar{\gamma}-\partial\bar{\gamma}
\bar{\partial} \gamma)+2 g_{ij}\partial 
X^i\bar{\partial} X^j\Big).}
We now perform a Euclidean rotation in both the worldsheet and
target space such that the Euclidean action is 
\eqn\euclidcont{S_0 = {1\over 4\pi
\ap}\int d^2z\Big(\partial \gamma
\bar{\partial}\bar{\gamma}+\partial\bar{\gamma}\bar{\partial} \gamma
- 2\pi \ap B (\partial \gamma
\bar{\partial}\bar{\gamma}-\partial\bar{\gamma}
\bar{\partial} \gamma)+2 g_{ij}\partial 
X^i\bar{\partial} X^j\Big).}
Note that, unlike the case of a $B$-field with only space-like
components, there is no factor of $i$ in the term in the Euclidean
action depending on $B$.

In order to obtain a finite worldsheet description in the NRCS limit, it is
convenient to introduce Lagrange multipliers $\beta$ and $\bar{\beta}$. 
 In
these variables the worldsheet theory \euclidcont\  is given by
\eqn\lagrangian{
 S_1 =\int \hskip-2pt {d^2z
\over 2\pi}\left(\beta \bar{\partial}
 \gamma + \bar{\beta}   
 \partial \bar{\gamma} - {2\ap\over 1+2\pi \ap
B} \beta\bar{\beta}+{1-2\pi \ap B\over 2 \ap}\partial
\gamma \bar{\partial} \bar{\gamma}+{1\over  \ap}g_{ij}\partial
X^i\bar{\partial} X^j\right),}
where \euclidcont\ is reproduced by integrating out $\beta$ and $\bar{\beta}$.
Therefore, in the strict decoupling limit \critical , one has
the following Lagrangian description of NRCS
\eqn\lagrangian{
S_1=\int {d^2z\over 2\pi}
\left(\beta\bar{\partial}\gamma
+\bar{\beta}\partial
\bar{\gamma}+{1\over 4 
\alpha_{eff}'}\partial \gamma \bar{\partial}\bar{\gamma}
+{1\over  \alpha_{eff}'}\partial X^i\bar{\partial} X_i\right).}
We note that the worldsheet theory of NRCS is defined in terms of more
variables than a conventional critical string theory since we have extra
$\beta$ and $\bar{\beta}$ variables. However, the CFT defined by
\lagrangian\ has
the correct Virasoro central charge to define a consistent string
action (see next subsection for more details). It is interesting to
note that the Lagrangian \lagrangian\ is invariant under the Galilean
group in the transverse coordinates. This is consistent with the
non-relativistic spectrum that we will find for NRCS.
It is crucial, for this symmetry to be realized,  that the 
description has the extra variables $\beta$ and $\bar{\beta}$.

We will concentrate on the conformal field theory of $\beta, \bar{\beta}$ and 
$\gamma, \bar{\gamma}$ since
the transverse coordinates lead to familiar contributions. The
equations of motion force $\beta$ and $\gamma$ to be holomorphic
 and $\bar{\beta}$ and $\bar{\gamma}$ to
be antiholomorphic. The Lagrange multiplier $\beta$
forces $\gamma$ to be a holomorphic map from the
worldsheet to the $(1+1)$-dimensional part of the target
space parametrized by coordinates $X^0$ and $X^1$. Therefore, it describes a
worldsheet instanton and the third term in the 
Lagrangian \lagrangian\  ${1\over 8 \pi \alpha_{eff}'} \partial \gamma
\bar{\partial} 
\bar{\gamma}$ is the instanton action.
We will show in the rest of this section that this Lagrangian 
reproduces the spectrum in \windingenergyformula . Moreover, we will
see that if the string 
worldsheet has a boundary that \lagrangian\ reproduces the open string
spectrum of NCOS and string interactions. In our formalism, one can
show that the decoupling of the massless open string modes exhibited
by \km\ at the disk level can be extended to all orders in
perturbation theory. Thus, \lagrangian\ can be
used to perform manifestly finite worldsheet computations for NCOS theories.
 In the later sections, we will use this Lagrangian to
describe amplitudes involving closed strings and higher loops.

\subsec{Closed string spectrum}

Here we consider a worldsheet without boundaries.
The equations of motion of \lagrangian\ imply that  $\beta(z)$ and $\gamma(z)$
are holomorphic and that $\bar{\beta}(\bar{z})$ and
$\bar{\gamma}(\bar{z})$ are antiholomorphic. Their OPE's are given by   
\eqn\opeclosed{
 \eqalign{ &\beta(z) \gamma(w) \sim {-1 \over z-w}~,~~~
           \bar{\beta}(\bar{z}) \bar{\gamma}(\bar{w}) \sim {-1
           \over \bar{z}-\bar{w}}\cr 
&\gamma(z)\beta(w) \sim {1\over z-w}~,~~~
           \bar{\gamma}(\bar{z}) \bar{\beta}(\bar{w}) \sim {1
           \over \bar{z}-\bar{w}}\cr 
&\gamma(z)\bar{\gamma}(\bar{w}) \sim  {\rm regular}~,~~~
  \beta(z) \bar{\beta}(w) \sim {-\pi  \over 2\alpha_{eff}'} 
\delta^{(2)}(z-w).}}
The variables $\beta$ and $\gamma$ behave analogously to the bosonic
ghost system, except 
for the contact term in $\beta(z)\bar{\beta}(\bar{w})$. The
conformal dimension of $\gamma$ is $(0,0)$  while the conformal dimension of
$\beta$ is $(1,0)$. Moreover, their contribution to the Virasoro central charge
takes the required value $c=2$. Note that although it seems that we
have added more degrees of freedom to the description, the worldsheet
degrees of freedom are identical to those of  two worldsheet scalars.
A similar story holds for the
$\bar{\beta}$ and $\bar{\gamma}$ 
system.

We will first consider the case when the $x^1$ coordinate is
noncompact. Then we can expand the
operators as
\eqn\oscillators{\eqalign{
  \gamma(z) & = \sum_{n=-\infty}^\infty \gamma_n z^{-n} ,\cr
  \beta(z) & = \sum_{n=-\infty}^\infty \beta_n z^{-n-1}}.}
Since $\gamma$ is holomorphic and $x^1$
is noncompact the standard space-time momentum term in $\gamma$ is not
allowed. Otherwise, $\gamma$ would be multi-valued function of
$\sigma$, which is only possible if the string is winding.
The oscillator modes satisfy the following commutation relation
\eqn\commutator{ 
[ \gamma_n, \beta_m ] = \delta_{n+m,0}.}

We now study the Virasoro constraints for NRCS. The energy momentum
tensor is given by 
\eqn\emtensor{
 T(z) = -\beta \partial \gamma ,} 
and the 
Virasoro generators are
\eqn\virflat{ L_n =\sum_m m \beta_{n-m} \gamma_m.}
In particular $L_0$ is the excitation level of
the $(\beta,\gamma)$ system, whose spectrum is positive
definite. Therefore  the Virasoro constraint
$L_0 + \tilde{L}_0 =1$ has no solutions
(except for the tachyon,
which is projected out in supersymmetric theories), 
where $\tilde{L}_0$ is the Virasoro generator for the rest 
of the system, which we assume
to be positive definite. Thus, the closed string has no physical
states. This can be easily understood. If all coordinates are
non-compact the background NS-NS B field can be gauged away without
changing the Hamiltonian of the theory and the closed string spectrum
is the usual one. Therefore, in the NRCS limit \critical ,  all the
closed strings 
acquire infinite proper energy and thus are unphysical.

When the $x^1$ direction is compactified on a circle
of radius $R$, there is a non-zero winding sector in $\gamma(z)$. The
mode expansion now reads
\eqn\winding{
\gamma(z) =  iwR\log z + \sum_{n=-\infty}^\infty \gamma_n z^{-n},}
and the Virasoro generator is then given by
\eqn\vircircle{
 L_n = -i\beta_n wR + \sum_m m \beta_{n-m} \gamma_m.}
Now the Virasoro constraint $L_0 + \tilde{L}_0 =1$ has a solution. 
As we will show later, all physical states are in
the vacuum of the $(\beta,\gamma)$-system. Thus the solution
to the Virasoro constraint is of the form
\eqn\asolution{
 i\beta_0 ={N \over wR}+{\alpha_{eff}'k^2\over 4wR},}
where $N$ is the conformal weight of the rest of the system
($N$ includes $-1$ for the bosonic string and
$-{1\over 2}$ for the NS sector of superstring)
and $k$ is the transverse momentum of the string.
According to the Lagrangian \lagrangian,
the canonical momentum $P$ conjugate to $\gamma$ is
not equal to $\beta$, but in the winding sector it is shifted by an
amount proportional to $\bar{\partial}\bar{\gamma}$ as in
\eqn\momentumconj{\eqalign{
{1\over 2}(P_0+P_1) &= {\beta\over 2\pi} - {1 \over 8\pi\alpha_{eff}'}
      \bar{\partial} \bar{\gamma} \cr
{1\over 2}(P_0 -P_1) &=  {\bar{\beta}\over 2\pi} - {1 \over
      8\pi\alpha_{eff}'} 
      \partial \gamma .}}
Taking the zero mode parts of these equations, one finds that
the total energy
$p_0$ and momentum $p_1$ of the closed string are given by
\eqn\energymomentum{\eqalign{
 {1\over 2}(p_0 + p_1) &= i\beta_0 + {1\over 4\alpha_{eff}'} wR,\cr
  {1\over 2}(p_0-p_1) &=  i\bar{\beta}_0 + {1\over 4\alpha_{eff}'} wR.}
}
Since $x^1$ is periodic, its conjugate momentum is quantized as
\eqn\momenta{ p_1 = {n \over R}.}
Thus we find
\eqn\solution{
p_0 = 
 {wR \over 2\alpha'_{eff}} + {\alpha_{eff}'k^2\over 2wR}+ 
{N + \bar{N}\over wR},}
with the level matching condition $N - \bar{N}= wn $.
In this way, we have recovered the non-relativistic spectrum in
 \windingenergyformula .

It is straightforward to prove the no-ghost theorem
in this case. We introduce the $(b,c)$ ghost and
write the BRST operator $Q_{BRST}$ as
\eqn\brst{ Q_{BRST} = Q_{-1} + Q_0,}
where
\eqn\qminus{ Q_{-1} = wR \sum_n \beta_{-n} c_n,}
and $Q_0$ is defined as the remainder. 
They obey
\eqn\anticomm{ Q_{-1}^2 = Q_0^2 =\{ Q_{-1}, Q_0 \} = 0.} 
Following the use of the BRST operator in
\ref\katoogawa{M. Kato and K. Ogawa, ``Covariant Quantization
of String Based on BRS Invariance,'' Nucl. Phys. B212 (1983) 443.}
(see also section 4.4 of \ref\bigbook{J. Polchinski, {\it String
Theory, Volume I} (Cambridge University Press, 1998).}),
one can show that the cohomology of $Q_{BRST}$ is 
isomorphic to that of $Q_{-1}$. The cohomology of
the quadratic operator $Q_{-1}$ is easy to evaluate
\ref\kugoojima{T. Kugo and I. Ojima, 
``Manifestly Covariant Canonical Formalism of the Yang-Mills
Field Theories I.'' Prog. Theor. Phys. 60 (1978) 1869.}
and it is spanned by 
the vacuum state $|0 \rangle_{\beta\gamma}
\otimes |0 \rangle_{bc}$ of $(\beta,\gamma)$ and
$(b,c)$, times any states in the rest of the system.
Assuming that the rest of the system is unitary,
this proves the no-ghost theorem of NRCS.

\subsec{Open string spectrum}

It is possible to consider D-branes in NRCS,
and what one gets is of course NCOS. Here
we show that the open string spectrum of NCOS
is reproduced in this way. Suppose the worldsheet is the
upper half plane with the boundary 
located at $\sigma^1=0$. In the bulk, $\gamma$ and $\beta$ are
still holomorphic. The equation of motion at the
boundary gives
\eqn\boundaryeq{
 \delta \gamma
\left( \beta + {1\over 4\alpha_{eff}'} \bar{\partial} 
\bar{\gamma}\right) = 0, ~~~
  \delta \bar{\gamma} \left( \bar{\beta} + {1\over 4\alpha_{eff}'}
  \partial \gamma \right) = 0.}
For a Dp-brane with $p \geq 1$, the boundary values of
$(\gamma,\bar{\gamma})$ are not fixed. Therefore 
\eqn\pboundary{
 \beta = - {1\over 4\alpha_{eff}'} \bar{\partial} \bar{\gamma}~,~~
   \bar{\beta} = -{1\over 4\alpha_{eff}'} \partial \gamma,}
at the boundary. This suggests that we analytically continue
$(\gamma,\beta)$ to $\sigma^1< 0$ and 
use \pboundary\ to identify $(\gamma,\beta)$ on $\sigma^1 \geq 0$
to $(\bar{\gamma},\bar{\beta})$ in $\sigma^1 \leq 0$.
Since
\eqn\opebulk{ \beta(z) \gamma(w) \sim -{1 \over z-w}~,~~
  \bar{\beta}(\bar{z}) \bar{\gamma}(\bar{w}) \sim
  -{1 \over \bar{z} - \bar{w}},}
we find
\eqn\opeboundary{ \gamma(z) \bar{\gamma}(\bar{w}) \sim
4\alpha_{eff}'\log (z-\bar{w}).}
For points on the real axis this reproduces the correct propagator for
open strings in NCOS. 

To compute the open string spectrum, we 
consider a worldsheet that is a strip $-\pi \leq \sigma^1 \leq \pi$, 
we identify $(\gamma,\beta)$ on $0\leq \sigma^1 \leq \pi$
to $(\bar{\gamma},\bar{\beta})$ on $-\pi \leq \sigma^1 \leq 0$
according to \pboundary . Thus we have the expansion
of these fields as
\eqn\stripexpansion{
  \eqalign{ \beta & = {1\over 2 \sqrt{\alpha_{eff}'}}\sum_n \bar{\alpha}_n
  \bar{z}^{-n-1}~,~~   
  \gamma =  x +  4 i\;p\;\alpha_{eff}'\log z +2
  \sqrt{\alpha_{eff}'}\sum_{n\neq 0} {\alpha_n \over n}z^{-n},  \cr
 \bar{\beta} & = {1\over 2 \sqrt{\alpha_{eff}'}}\sum_n \alpha_n z^{-n-1}~,~~
  \bar{\gamma}=  \bar{x} +4 i\;\bar{p}\;\alpha_{eff}'\log \bar{z} +2
  \sqrt{\alpha_{eff}'}\sum_{n\neq 0} {\bar{\alpha}_n \over
  n}\bar{z}^{-n},}}
and non-zero commutators are
\eqn\standardcommutator{
  [\alpha_n , \bar{\alpha}_m] =n \delta_{n+m,0},~~
[x,\bar{p}]=[\bar{x}, p] = i.}
Note that, unlike the case of closed string, the space-time
momentum term $4ip\;\alpha_{eff}' \log z$ is allowed in
$\gamma$ even when $x^1$ is non-compact. This is because
we can choose the branch cut of $\log z$ to be away from 
the worldsheet.  The Virasoro generators are then
\eqn\openvirasoro{
  L_n = \sum_m \bar{\alpha}_{n-m} \alpha_{m}.}
Thus we reproduce the standard open string spectrum.

\subsec{Free the $U(1)$ to all order}

In \km, it was shown that any tree level amplitude containing
a massless open string state in NCOS$_2$ vanishes when the longitudinal
direction is non-compact. This is what is expected from the point
of view of the S-dual theory, where the massless open
strings correspond to the free $U(1)$ gauge fields and their
superpartners in the $(1+1)$-dimensional $U(N)$ gauge theory. 
Using the formalism developed here, it is straightforward to generalize this
result to all orders in perturbation theory. 
The open string amplitude
on a worldsheet with $h$ holes and $g$ handles  
is computed by considering a closed Riemann surface of genus
$2g$ with a complex conjugate involution such that the fixed 
point set of the involution gives the boundaries of the
open string worldsheet.
Since there is no vertex operator inserted away from the boundary
(when the longitudinal direction is non-compact there is no
closed string physical state),  $\gamma(z)$ is holomorphic everywhere
except at the boundaries. Moreover, the vertex operator for a massless
open string state is also holomorphic (vertex operators for massive
states are not holomorphic, they also depend on $\bar{\gamma}$).
Since the sum of the boundaries of the $h$ holes obtained as the
fixed point set of the involution is homologically trivial
on the genus $2g$ surface, the contour integral of
the vertex operator can be deformed away 
through the middle of the Riemann surface. This proves
the decoupling of the massless open string states
to all order in the perturbation theory.

\newsec{Tree amplitudes}

In this section we will compute the scattering
amplitude of four physical closed string states and show
that it factorizes properly into non-relativistic closed 
string poles. Moreover, we will see that the truncated 
closed string scattering amplitudes  have a
different high energy behavior than in conventional string theory.

For simplicity, we will compute the $4$-tachyon amplitude. 
Since all the physical states are 
in the vacuum of the $(\beta,\gamma)$-system as we saw
in the last section,  the
essential novelty of NRCS is captured by
the tachyon amplitude. The vertex operator 
for a closed string tachyon is
given by\foot{We will not include the cocycles which only change relative
signs between amplitudes. The factor $\sqrt{w}$ is
included for later convenience.}
\eqn\vert{
V(v,\bar{v},k;z,\bar{z})=g \sqrt{w}:e^{i\bar{v} \gamma(z) 
+ i wR \int^z \beta + iv \bar{\gamma}(\bar{z})  
+ i wR \int^{\bar{z}} \bar{\beta} 
+ ik\cdot X(z,\bar{z})}:.}
Since
\eqn\momope{ \eqalign{
 \gamma(z) V(z') & \sim i wR \log(z-z') V(z'),\cr
  \beta(z) V(z') &\sim {-i\bar{v} \over z-z'} V(z'),  }}
the vertex operator carries $(\beta_0,\bar{\beta}_0)$
eigenvalues of $(-i\bar{v},-iv)$ and winding number $w$. 
Thus, according to 
\energymomentum , the energy $\epsilon$ and 
the longitudinal momentum $n/R$ of the tachyon state are given by
\eqn\mome{\eqalign{ 
\epsilon&= v+\bar{v}+{1\over 2\alpha_{eff}'} wR \cr
{n \over R}&= {\bar{v}-v}.}}
Let us evaluate the 4-point amplitude on the sphere,
 $\langle V_1(z_1) V_2(z_2) V_3(z_3) V_4(z_4) \rangle$,
by performing the functional integral with the Lagrangian
\lagrangian. The extremum of the functional integral is
given by 
\eqn\sphereextr{
  \eqalign{\gamma(z) & = i\sum_{a=1}^4 w_a R\log (z-z_a), \cr
            \beta(z) & =-i\sum_{a=1}^4 {\bar{v}_a \over z-z_a}.}}
For closed string amplitudes, winding number has to be 
conserved\foot{The winding number is not
conserved when one considers worldsheet with boundaries.}, 
such that $\sum_a w_a = 0$.
Since the action is free, we can evaluate the amplitude by
substituting \sphereextr\ back into the integrand\foot{As is
always the case with the Gaussian integral, the same result
is obtained by substituting the extremal value \sphereextr\ 
into the product of the vertex operators $V_1V_2V_3V_4$ alone
and by taking its square root.}. The amplitude is 
            given by
\eqn\spherefourresult{  \eqalign{&
 \langle V_1(z_1) V_2(z_2) V_3(z_3) V_4(z_4) \rangle = \cr
& = \sqrt{w_1\cdots w_4}\prod_{a \neq b} (z_a - z_b)^{-\bar{v}_aw_bR}
(\bar{z}_a - \bar{z}_b
   )^{-v_a w_bR}|z_a-z_b|^{{\alpha_{eff}'\over 2} k_ak_b}\cr
& = \sqrt{w_1 \cdots w_4}
\prod_{a < b} |z_a - z_b|^{-(\epsilon_a + \epsilon_b)(w_a + w_b)R
 + {R^2 \over 2\alpha_{eff}'}(w_a+w_b)^2  + 
{\alpha_{eff}'\over 2} (k_a+k_b)^2-4}
.} }
Here we used the on-shell condition for the tachyon, 
\eqn\tachyononshell{\eqalign{
\epsilon_a &= {w_a R \over 2\alpha'_{eff}}
+ {\alpha_{eff}' k_a^2\over 2w_aR} - {2 \over w_aR}~, \cr
v_a& = \bar{v}_a~,~~~~~~~~~~(a=1,\cdots,4).}}
It is a good test of our formalism to compute 
the same correlation function using the standard 
closed string theory and then take the NCOS limit 
(3.2). One can verify that \spherefourresult\ is reproduced
in the limit.  
The tachyon scattering amplitude is then given by
\eqn\ampli{{\cal A} = ig^4 C_{sphere}
\int d^2 z 
\langle V_1(0) V_2(z) V_3(1) V_4(\infty) \rangle .}
Here $g$ is the closed string coupling constant and
$C_{sphere}$ is the normalization constant that normalizes the path
integral of the string when the topology of the worldsheet is the
sphere. The normalization constant can be found by unitarity. Namely,
the amplitude in \ampli\ has poles associated with intermediate closed
string states and a straightforward application of the optical theorem
determines it. Therefore, by repeating the analysis in, for example,
section 6.6  of \bigbook , we find 
\eqn\spherenormalization{C_{sphere} = {2 \pi \over g^2 R}.}
This means that even though the theory is defined in the
$g_s\rightarrow \infty$ limit, that the closed string theory has a
sensible perturbation expansion in powers of $g$.
This is consistent with the observation made in [4] regarding
closed string loop diagrams in NCOS.

The amplitude given by \spherefourresult\ and
\ampli\ is very similar to the familiar Virasoro-Shapiro
amplitude. It has poles in the energies in the intermediate
channels, and they are located at
\eqn\vira{\eqalign{
\epsilon_a + \epsilon_b
 &= {(w_a+w_b)
R \over 2 \alpha_{eff}'}+{ \alpha_{eff}' (k_a+k_b)^2\over 2(w_a+w_b)R}
+{2n-2\over (w_a+w_b)R},\cr
& ~~~~~~~~~~~~~~(n=0,1,2,\cdots).}}
This is precisely the closed string spectrum of NRCS, as required by
unitarity. From \ampli\ we can also see that NRCS exhibits a
different behaviour of high-energy, fixed-angle scattering
amplitudes. Since the mass-shell condition of the strings is
non-relativistic, the dependence of the amplitude on energy $E$ is ${\cal
A}\sim e^{-E}$, as opposed to the conventional dependence ${\cal
A}\sim e^{-E^2}$.

Although there are no physical states in the sector
with 0-winding number and in particular no graviton
in the spectrum, there is an instantaneous Newtonian
potential between winding strings. To see this, 
let us consider the process in which the winding
number is not exchanged among strings, $i.e.$ $w_1+w_3=0$
and $w_2+w_4=0$. In this case, the correlation
function \spherefourresult\ becomes
\eqn\colliding{\eqalign{
& \langle V_1(z_1) V_2(z_2) V_3(z_3) V_4(z_4) \rangle 
= w_1 w_2 \Big(|z_1-z_3||z_2-z_4|\Big)^{\alpha_{eff}' (k_1+k_3)^2 - 4}
\times \cr
&~~~~~~~\times \Big(|z_1-z_2||z_3-z_4|\Big)^{
-(\epsilon_1+\epsilon_2)(w_1+w_2)R + {R^2 \over 2\alpha_{eff}'}(w_1+w_2)^2
 +{\alpha_{eff}' \over 2}(k_1+k_2)^2 - 4} \times \cr
&~~~~~~~\times \Big(|z_1-z_4||z_2-z_3|\Big)^{
-(\epsilon_1+\epsilon_4)(w_1-w_2)R + {R^2 \over 2 \alpha_{eff}'}(w_1-w_2)^2
 + {\alpha_{eff}'\over 2}(k_1+k_4)^2-4}.}}
Since the winding number along the $(k_1+k_3)$-channel is $0$, 
no physical states are propagating in this channel. 
Nevertheless, after doing the $z$-integral in \ampli,  one finds
that there are contributions from
exchange of off-shell states in the 0-winding number sector.
In particular,  the leading long-range contribution to
the 4-point amplitude contains $\sim (k_1+k_3)^{-2}$, 
corresponding to the Newtonian potential. 
Thus, even though the theory contains no gravitons,
there is an instantaneous gravitational force between
winding strings.  

\newsec{Loop Amplitudes}

In this section, we will compute
the one-loop free energy at finite temperature and one-loop 
corrections to $N$-point functions of winding states.
We will also examine the general structure of higher loop
amplitudes and demonstrate that there is a sensible
perturbative expansion of NRCS.
 
On a genus-$g$ surface, the $\beta$-field (the $(1,0)$-form
we introduced in section 3 as a Lagrangian multiplier) has
$g$ zero modes. If we were quantizing the bosonic ghost system,
we would introduce delta-functions in the path integral to absorb
these zero modes.  
However, one can show that the rules of the NRCS perturbation theory
deduced from the factorization conditions do not call for
these delta-functions. Thus one may naively think that zero mode integrals 
 are divergent in NRCS. This would be 
similar to the problem in DLCQ of field 
theory\lref\my{T. Maskawa and K. Yamawaki,
``The Problem of $P^+=0$ Mode in the Null Plane
Field Theory and Dirac's Method of Quantization,''
Prog. Theor. Phys. 56 (1976) 270.}
\lref\bp{S. Brodsky and C. Pauli, ``Discretized Light Cone Quantization:
Solution to a Field Theory in One Space One Time Dimensions,''
Phys. Rev. D32 (1985) 2001.} \refs{\my,\bp}, where integrals 
over states carrying zero longitudinal momentum pose difficulties 
in evaluating loop amplitudes \ref\dlcq{S. Hellerman and J. Polchinski,
``Compactification in the Lightlike Limit,''
Phys. Rev. D59 (1999) 125002, {\tt hep-th/9711037}.}.

It turns out that, whenever we evaluate physical 
observables such as the temperature dependent part of the free energy 
and scattering amplitudes of closed strings with non-zero winding 
numbers, the amplitudes contain terms of {\it stringy nature} 
which depend on all the $g$ zero modes of $\beta$, so that  
the zero-mode integrals are convergent.
It is easy to understand where these terms come from; 
they appear because $\beta$ is a Lagrange multiplier
which constrains $\gamma$ to be a holomorphic map 
from the worldsheet to the $(1+1)$-dimensional part of the
target space. If vertex operators for winding states are
inserted on the worldsheet, a holomorphic map $\gamma$, if it
exists, has to be a 
non-trivial one since the image of the worldsheet 
has to wind around each of the vertex operators. As
we will show below, a non-trivial 
holomorphic map from the
worldsheet to the $(1+1)$-dimensional part of the
target space, which is a cylinder,
exists only in a subspace of codimension $g$ 
of the moduli space of a genus $g$ Rieman surface. 
The integral over the $g$ zero modes of $\beta$
gives a delta-function which exactly picks up the subspace 
where the holomorphic maps exist.  

On the other hand, if we consider amplitudes which do
not contain winding strings, such as the vacuum amplitude
at zero temperature, then the zero-mode integral gives a divergence.
In this case, $\gamma$ can be a {\it trivial} map which maps
the worldsheet to a point in the target space. Such a map
exists everywhere on the moduli space of the worldsheet,
and therefore the worldsheet amplitude is independent of 
the $g$  zero modes of $\beta$. The integral over these
zero modes is then flatly divergent.   
If one traces through the NCOS limit in section $3$, one
finds that it is exactly the type of divergence that was
pointed out in \dlcq. Fortunately all the physical states in 
NRCS have non-zero winding number, and 
these divergent amplitudes have no physical meaning
and can be safely ignored.

We will demonstrate these points by computing one-loop 
amplitudes and 
show how they are generalized to higher loops.

\subsec{Free energy}

The one-loop free energy at temperature $T$ is evaluated by
performing an Euclidean rotation of the target space-time coordinate   
and periodically identifying $X^0 \sim X^0 + T^{-1}$. The path integral
then involves a sum over maps $(\gamma,\bar{\gamma})$
from the worldsheet torus of modulus $\tau$ to the target space
torus of periods $(T^{-1}, 2\pi R)$. 

The zero mode dependence of the free energy can be computed by
performing the path integral over the maps from the worldsheet to
space-time.  Thus we write
\eqn\windinggamma{
 \gamma = \left(2\pi n R  + i {m \over T} \right) {\sigma^0 \over 2\pi} + 
          \left(2\pi w R + i {s \over T} \right){\sigma^1\over 2\pi} +
         ({\rm periodic}),}
where $0 \leq \sigma^0, \sigma^1 < 2\pi$ and
$(n,m,w,s)$ are integers labeling the different winding sectors. For
 this $\gamma$,  
\eqn\delbargamma{
\bar{\partial} \gamma = {i \over 4\pi Im \tau}
 \left[ 2\pi nR + i {m \over T}
 -  \tau \left( 2\pi wR + i {s \over T} \right) \right]
 + \bar{\partial} ({\rm periodic}).}
On the other hand, $\beta$ can be written as
$\beta = \beta_0 + \partial ({\rm periodic}),$ where
$\beta_0$ is the zero mode. The worldsheet action depends
on $\beta_0$ as
\eqn\instantonactionforfreeenergy{
 S = i \beta_0 \left[2\pi  nR + i {m \over T}
 -  \tau \left( 2\pi wR + i {s \over T} \right) \right]
  + \cdots.}
Thus the integral over $\beta_0$ gives a delta-function which
fixes the worldsheet modulus $\tau$ at
\eqn\taufixed{
 \tau ={2\pi nR + i m{1 \over T} \over 2\pi wR + i s{1 \over T} }.}
Thus the $\tau$-integral becomes a sum over these special
points on the worldsheet moduli space. These are the points
at which there are holomorphic maps from the worldsheet to
the target space. 

The one-loop free energy is obtained by a sum over the
integers $(n,m,w,s)$ such that $\tau$ is in the fundamental
domain of the moduli space. To do the summation, it is
convenient to use the trick invented in 
\ref\joetrick{J. Polchinski, ``Evaluation of the One Loop 
String Path Integral,'' Commun. Math. Phys. 104 (1986) 37.}
to trade the sum over $s$ for the sum over copies of the fundamental
domain. If $(m,s)\neq (0,0)$, there is an $SL(2,Z)$ transformation
which sends $(m,s)$ to $(m,0)$ with $m > 0$, and it also maps the fundamental
domain of $\tau$ into the strip, $|{\rm Re}~ \tau| \leq 1/2$,
in the upper half-plane ${\rm Im}~\tau \geq 0$. The sum over $s$ covers
the strip exactly once by copies of the fundamental domain. On the other
hand, the $(m,s) = (0,0)$ term is
independent of the temperature $T$ and corresponds to the zero
temperature vacuum energy. We will ignore this contribution since
it has no physical meanings in NRCS and it vanishes 
in supersymmetric theories anyway. Thus we have 
\eqn\tauwiththreeintegers{
  \tau = {{2\pi nR + i m{1 \over T} \over 2\pi wR} },}
and we sum over integers $(n,m,w)$. Since $m > 0$ and
$\tau$ must be in the strip in the upper half-plane,
we require $w> 0$ and $|n| \leq w/2$ ($n$ at the boundary
$n=\pm w/2$ is counted with a factor $1/2$). 

We can now evaluate the path integral over $\gamma$ and
$\bar{\gamma}$. The zero mode piece is obtained by evaluating the
instanton action. Therefore, substituting \windinggamma\ (with $s=0$)
into the action \lagrangian\ and  
evaluating it at the points \tauwiththreeintegers\ of the moduli
space, we find that the zero mode part of the action is  
\eqn\instantonactionforfreenergy{
\eqalign{ S &= \int d^2z \left( {1\over 8\pi \alpha_{eff}'}
 \partial \gamma \bar{\partial}\bar{\gamma} + \cdots \right) \cr
& = {mwR \over 2 \alpha_{eff}' T} + \cdots .}}
As usual, the contribution from the non-zero modes of $(\beta,\gamma)$ 
is canceled by the determinant of the $(b,c)$ ghost system.
Therefore, the one-loop contribution to the free energy takes the form
\eqn\onelooppartition{
 F(T) = -\sum_{n,m,w} {T \over wm}
 \sum_{h,\bar{h}} D(h,\bar{h}) \exp \left[ 
 -{m \over T} \left( {wR \over 2\alpha_{eff}'} 
+ {h + \bar{h} \over wR}\right)
         + 2\pi i {n\over w} (h - \bar{h}) \right] .}
This is obtained by evaluating the partition function 
of the worldsheet theory at the special points 
\tauwiththreeintegers\ on the moduli space. Here 
$(h,\bar{h})$ are the conformal weights coming from
the transverse excitations of the string 
and $D(h,\bar{h})$ is their multiplicity.
To simplify the expression in \onelooppartition, 
we have included in $h$ the contribution from the  
transverse momenta $k$. Thus, in comparison with
the notion in section 3, $h$ and $\bar{h}$ are defined as
\eqn\confweight{
 h = {\alpha_{eff}' \over 4}\, k^2 + N~,~~~
 \bar{h} = {\alpha_{eff}' \over 4}\, k^2 + \bar{N}.}

The factor $-T/wm$ in  \onelooppartition\ is determined as follows.
The $\beta_0$ integral with the action \instantonactionforfreeenergy\
gives a factor $(2\pi w R)^{-2}$ times the delta-function for
$\tau$ (we set $s=0$ in \instantonactionforfreeenergy). The measure
for the $\tau$-integral contains the factor   
\eqn\taumeasure{ {1 \over {\rm Im}~\tau} = 
{2\pi w R T\over m}.}
The zero mode integral of $\gamma$ gives the volume $2\pi R/T$ of
the target space. Finally, the definition of the one-loop free energy
is $F = - T Z(T)$, where $Z(T)$ is the one-loop vacuum amplitude
at temperature $T$. Combining these factors together, we obtain
\eqn\factors{
  {1 \over (2\pi w R)^2} {2 \pi wR T \over m} {2\pi R \over T}
  (-T) = - {T \over wm},}
as in  \onelooppartition. 

The sum over $n$ in $|n| \leq w/2$
gives the constraint $h - \bar{h} \equiv 0$ (mod~$w$), which
we recognize as the level matching condition. 
After summing over $m$, the free energy given by
\onelooppartition\ becomes
\eqn\freenergyresult{
 F(T) = T \sum_{w=1}^\infty \sum_{h, \bar{h}} D(h,\bar{h})
\log\left( 1 -e^{-{E(w,h,\bar{h})\over T}} \right).}
This is the conventional expression for the one-loop free 
energy of quantum field theory. Here
\eqn\energyagain{
 E(w,h,\bar{h}) = {wR \over 2 \alpha_{eff}'}
 + {h + \bar{h} \over wR} = {wR \over 2 \alpha_{eff}'}
 + {\alpha_{eff}'  k^2 \over 2wR} +{N + \bar{N} \over wR}.}
With the level matching condition for $(h,\bar{h}$),
the expression for $E(w,h,\bar{h})$ precisely agrees the  
energy spectrum of closed strings in NRCS computed in 
section 3, with the correct multiplicity factor. 

The computation of the free energy described here
is similar to the one for string in $AdS_3$ discussed
in \mos.
In that case, the one-loop amplitude has poles exactly
at the special points in \taufixed ,  
and they are due to the presence of long strings
winding near the boundary of $AdS_3$. Here we have delta-functions 
at these points and they correspond to the closed
strings winding in the $x^1$ direction. 

We now study the high temperature behaviour of the free energy.
It is clear that, for a fixed  winding number $w$,
that the free energy is convergent for any temperature $T$. 
However, there might be a divergence when one performs the sum over
winding modes. To see whether the sum over $w$ gives a divergence, 
we use the Cardy's formula
\eqn\modular{
  \sum_{h,\bar{h}} D(h,\bar{h}) 
\exp\left(2\pi i\tau h - 2\pi i \bar{\tau} \bar{h}\right)
 \sim \exp\left[ {2\pi ic \over 24}\left({1 \over \tau}
 - {1 \over \bar{\tau}} \right)\right],}
for ${\rm Im}~\tau \rightarrow 0$. Here $c$ is the central charge
of the rest of the system, and $c={12}$ for Type II superstring. 
Therefore, the high temperature behaviour of NRCS is given by
\eqn\highT{\eqalign{& \sum_{|n| \leq w/2}
 \sum_{h,\bar{h}} D(h,\bar{h}) \exp \left[ 
 -{m \over T} \left( {wR \over 2\alpha_{eff}'} 
+ {h + \bar{h} \over wR}\right)
         + 2\pi i {n\over w} (h - \bar{h}) \right]\cr
& ~~~~~~~~\sim 
  \exp\left[  2\pi w R \left( {\pi c T \over 6 m}
                   - {m \over 4\pi \alpha_{eff}' T} \right)\right],}}
for large $w$. 
Therefore, the sum over winding states will be divergent whenever $T
\geq T_{H}$, where 
\eqn\hagedorn{
 T_{H} = {1 \over 2\pi} \sqrt{{6 \over  \alpha_{eff}'c}}.}
This gives the Hagedorn temperature of NRCS.
For Type II NRCS we find
that $T_{H}={1\over 2\pi \sqrt{2\alpha_{eff}'}}$.
It coincides with the location of the Hagedorn transition 
of NCOS studied in  \lref\ggkrw{S.S. Gubser, S. Gukov, I.R. Klebanov,
M. Rangamani, and E. Witten, ``The Hagedorn Transition in Non-Commutative
Open String Theory,'' {\tt hep-th/0009140}.}
\refs{\km,\ggkrw}. Just like in conventional
string theory, the Hagedorn temperature $T_{H}$ is the temperature
at which the tachyonic mode which appears in the spectrum
\energyagain\ due to  the finite temperature GSO projection becomes
massless. 
It would be interesting to study the behaviour of the closed
strings of NRCS above the Hagedorn temperature. In NRCS, 
the breakdown of the thermal ensemble 
may not occur unlike for 
relativistic closed string theories 
since there is no graviton and the Hamiltonian
is positive definite. However, there can be a Jeans instability.

\medskip
\noindent
$\underline{{\it Higher~loops}}$
\medskip

We have demonstrated explicitly that the $\beta$ zero mode
integral is convergent when one computes the one-loop free
energy. As a result of the $\beta$ zero mode integral, the integral over 
$\tau$ is localized to a sum over the 
 special points in the moduli space of the worldsheet torus 
where there is a holomorphic map from the worldsheet
to the target space torus. It is straightforward to generalize 
this observation to higher loops. A simple computation shows that a
map from a genus $g$ worldsheet to the target 
space torus exists only on a $(2g-3)$-dimensional subspace.
Such a holomorphic map exists whenever the following condition on the 
worldsheet period matrix $\Omega_{ij}$ ($i,j = 1, \cdots, g$)
is satisfied
\eqn\highergenera{
 G_i(\Omega)
=\sum_{j=1}^g  \Omega_{ij} \left(2\pi w^j R + i {s^j \over T} \right)
 + 2 \pi n_i R + i {m_i \over T}=0,}
for some integers $(n_i, m_i, w^i, s^i)$. 
This generalizes \taufixed\ for $g=1$. 
On a genus $g$ surface, $\beta$ has $g$ linearly independent
zero modes, and their integrals give delta-functions imposing
the constraints $G_i=0$ ($i=1,\cdots, g$). 

\subsec{$N$-point amplitudes}

Here we analyze the effects of the $\beta$ zero mode integrals on the
scattering amplitudes. We will discuss the $N$-point tachyon
amplitudes for simplicity, 
but a generalization to amplitudes involving arbitrary external
states is straightforward. 
As in the case of the tree amplitudes discussed in the 
last section, the computation of  $\langle \prod_{i=1}^n 
e^{i\bar{v}_a\gamma(u_a) + v_a\bar{\gamma}(\bar{x}_a)} \rangle$
requires finding $(\beta,\gamma)$ which are holomorphic away
from $u_a$'s and behave near $z=u_a$ as
\eqn\gammaontorus{\eqalign{
  \gamma(z) &\sim iw_a R \log(z-u_a),\cr
\beta(z)& \sim {-i\bar{v}_a \over z-u_a},~~~~~(z \rightarrow u_a),}}
where $w_a$ is the winding number for the $i$-th external state.
The functional integral is non-zero only when such $\beta$ and $\gamma$
exist. As for $\beta(z)$, there is always a meromophic one-form
given by
\eqn\alwaysbeta{ \beta(z) = -i\sum_{a=1}^N \bar{v}_a \partial_z \log
\vartheta_1(z-u_a) + {\rm const},}
where $\vartheta_1(z)$ is the elliptic theta function. This
$\beta$ is single-valued on the worldsheet torus as far as
 momentum is conserved, $\sum_a v_a = 0$. On the other hand,
$\gamma(z)$ obeying \gammaontorus\ does not always exist. To see
this, let us try
\eqn\trial{ \gamma(z) = i\sum_{a=1}^N w_a R \log\vartheta_1(z-u_a) + c z,}
for some constant $c$.  Due to the quasi-periodicity of the
elliptic function, we find
\eqn\gammaperiod{
\eqalign{ \gamma(z+2\pi) & = \gamma(z) + 2\pi c, \cr
\gamma(z + 2\pi \tau) & = \gamma(z) + 2\pi \left(-R \sum_{a=1}^N 
w_a u_a + c\tau
\right),}} 
where we assumed that the winding number is conserved,
$\sum_a w_a = 0$. By requiring that $\gamma(z)$ is periodic
modulo the target space periodicity $\gamma \sim \gamma + 2\pi R$, 
we find that $c$ must be of the form $c=mR$ for some
integer $m$ and 
\eqn\positions{
 \sum_{a=1}^N w_a u_a = n + m\tau,}
for some integer $n$. This gives a condition on the locations of
the $N$ points. 
Thus we find that, rather than being divergent, the 
integral over the zero mode of $\beta$ imposes the condition
\positions\ on the locations of the vertex operators on the worldsheet.

\medskip
\noindent
$\underline{\it Higher~loops}$
\medskip

It is straightforward to generalize this result to higher loops. 
On a genus $g$ worldsheet, the holomorphic map
$\gamma(z)$ winding $w_a$-times at $z=u_a$ should be
of the form,
\eqn\higherg{
 \gamma(z) = i\sum_{a=1}^N w_a R \log E(z,u_a)
           + \sum_{i=1}^g c^i \int^z  \omega_i,}
for some constants $c_i$, 
where $\omega_i$ are holomorphic one-forms, and
$E(z,w)$ is the prime form, a $(-{1\over 2})$-differential
with respect to $z$ and $w$ that vanishes linearly on the
diagonal $z=w$ only (see, for example, section IIIb.1 of
\ref\mumford{D. Mumford, {\it Tata Lectures on Theta II} 
(Birkh\"auser, 1984).}).
The periodicity of $\gamma$ in the $\alpha$-cycles of 
the worldsheet requires that $c^i$ must be of the form
$c^i = 2\pi m^i R$ for some integers $m^i$, and 
the periodicity around the $\beta$-cycles requires
\eqn\betaperiod{
 G_i = \sum_{a=1}^N w_a \int^{u_a} \omega_i -\left(
     n_i + \sum_{j=1}^g \Omega_{ij} m^j \right)= 0,}      
for some integers $n_i$. This imposes $g$ conditions on
the $(3g-3+N)$-dimensional moduli space of the genus $g$ 
worldsheet with $N$ points. To verify that 
the $g$ conditions $G_i=0$ are independent of each other
and pick up a codimension $g$ subspace, we need
to compute their partial derivatives with respect to
the worldsheet moduli $y_I$ ($I = 1, \cdots , 3g-3$)
and the vertex operator locations $u_a$. They are
given by
\eqn\derivatives{\eqalign{
  {\partial G_i \over \partial u_a}
 & =  \omega^i(u_a)w_a, \cr  
{\partial G_i \over \partial y_I}
 & = \int d^2z \omega_i(z)   \eta^I(z,\bar{z}) 
 \left( \sum_{a=1}^N w_a \partial_{z} \log E(z,u_a)
     - 2\pi i\sum_{j=1}^g m^j \omega_j(z) \right)
,}}
where $\eta^I$ are the Beltrami differentials
associated to the moduli $y_I$. Note that the partial derivatives
of $G_i$ are all of the form, $\int d^2 z\, \omega_i(z) \nu(z,\bar{z})$
for some differential $\nu$. Since 
\eqn\nodegenerate{
  {\rm det}_{i,j =1,\cdots g} \left( \omega_i(z_j) \right)
 \neq 0,}
for generic $g$ points $z_i$, it is clear that
the rank of the $(3g-3+N) \times g$ matrix
$(\partial_{u_a}G_i, \partial_{y_I} G_i)$
is generically $g$. Thus the integral over the  $g$ zero modes of $\beta$
exactly pick up the subspace of the moduli space where
the holomorphic map $\gamma(z)$ exist.

\newsec{Relation to DLCQ}

In this section we show that the NRCS limit we have studied is related by
T-duality to the DLCQ limit of string theory. This follows by
performing T-duality along the circle of radius $R$ where the B-field
lies. After T-duality, we get string theory without any background
B-field, with a metric
\eqn\tdualmetric{
           g_{\mu\nu} = \left( \matrix{ -1+(2\pi \alpha' B)^2
 & 2\pi \alpha' B \cr
 2\pi \alpha' B & 1 \cr}\right),}
and where the radius of the circle is $\alpha'/R$. In the NRCS limit
the metric is given by
\eqn\metbef{
ds^2=-{\ap\over \alpha_{eff}'} (dx^0)^2 + 2 dx^0dx^1+(dx^1)^2.}
We now rescale coordinates  such
that  $x^1 \rightarrow {\alpha'
\over \alpha'_{eff}} x^1$. In the NRCS ($\ap\rightarrow 0$) limit the
metric in these 
coordinates is 
\eqn\afterrescape{
{1 \over \alpha'}ds^2 =
      {1 \over \alpha_{eff}'}\big[
 - (dx^0)^2 + 2 dx^0 dx^1 \big], }
and the periodicity of the compact direction is given by 
\eqn\tdualperiodic{
x^1 \sim x^1 + 2 \pi{\alpha_{eff}' \over R}.} 
Since in this limit the $x^1$ coordinate is light-like, we see
that DLCQ of closed string theory with string scale $\alpha_{eff}'$ and
null radius $\alpha_{eff}'/R$ is T-dual to NRCS.

Because of the relation between NRCS and DLCQ, the formalism
developed in this paper gives a useful description of DLCQ closed
string theory also. In \lref\bilal{A. Bilal, ``A Comment
on Compactification of M Theory on an (almost) Lightlike
Circle,'' Nucl. Phys. B521 (1998) 202, {\tt hep-th/9801047};
``DLCQ of M Theory as the Lightlike Limit,''
Phys. Lett. B435 (1998) 312, {\tt hep-th/9805070}}
\lref\uehara{S. Uehara and S. Yamada,
``On the DLCQ as a Light-like Limit in String Theory,''
{\tt hep-th/0008146}.} \refs{\bilal,\uehara}, loop amplitudes of
DLCQ closed string theory were studied to address the issue
of divergence due to the longitudinal zero modes 
\refs{\my,\bp,\dlcq}. In particular, it was found in \bilal\
that one-loop scattering amplitudes, when external strings
carry non-zero longitudinal momenta, have finite DLCQ limits
and that the positions of the vertex operators
are constrained in the limit.
These constraints can be viewed as the T-dual of 
\positions\ in NRCS. The description of 
NRCS developed here does not involve the process of
taking the NCOS limit, and thus loop amplitudes
are manifestly finite. In fact, the one-loop observation in \bilal\
has a straightforward generalization to higher loops, as we saw
in \betaperiod . Evidently, unlike in ordinary field theories, 
the longitudinal zero modes do not cause
a problem in DLCQ closed string theory. In the case of
Type IIA string theory, this is related, via the hypothetical
11-dimensional Lorentz invariance of M theory, to the existence
of the smooth DLCQ limit of M Theory described by the finite $N$
Matrix Theory \lref\susskind{
L. Susskind, ``Another Conjecture about
M(atrix) Theory,'' {\tt hep-th/9704080}.}
\lref\senm{A. Sen, ``D0-branes on $T^n$ and Matix Theory,''
 Adv. Theor. Math. Phys. 2 (1998) 51, {\tt hep-th/9709220}.}
\lref\seim{N. Seiberg, ``Why is the Matrix Model Correct?''
Phys. Rev. Lett. 79 (1997) 3577, {\tt hep-th/9710009}.}
\refs{\susskind,\senm,\seim}.

For closed strings, the relation between NRCS and DLCQ
provides another 
way to understand the origin of the non-relativistic dispersion 
relation. The non-relativistic
limit describe in section $2$, however, is much broader
and includes cases that are not
related to DLCQ, as we will see in the next
section.

\newsec{Galilean Theories}

In this section we find new theories whose excitations satisfy a
non-relativistic dispersion relation. The light degrees of freedom
that survive the low energy limit are light-branes\foot{These theories
do not have background branes, unlike the theories discussed in \gmss .}.
Depending on which
background gauge field one tunes to its critical value, different
brane states are light while the rest of the states in M Theory decouple.

We will first consider M Theory limits where the light degrees of
freedom are membranes and fivebranes. We will call these theories GM
(Galilean membrane)
and  GF (Galilean five-brane) respectively. The first one is obtained
by tuning to the critical value the background three-form and the
second one by turning on the background six-form of M Theory. The low
energy limit is taken such that the terms in the worldvolume action
depending on the transverse coordinates to the background remain finite
in the limit. Just as for NRCS, the spectrum is modified
if the brane directions are compactified, otherwise there are no
finite energy physical excitations surviving the limit.

Therefore, the low energy limit leading to GM is given
by\foot{Throughout the rest of the paper we will have in mind
compactification on tori. It is straightforward to generalize the
decoupling limits when branes wrap curved geometries.}
\smallskip
\eqn\GMlim{\eqalign{
g_{\mu\nu}&=\eta_{\mu\nu} \quad \mu,\nu=0,1,2\cr
g_{ij}&={l^3_p\over l^3_{eff}}\delta_{ij} \quad i,j=3,\cdots,10\cr
C_{012}&=T_{M2}-T_{eff}},}
\smallskip
\noindent
with the eleven dimensional Planck scale $l_p\rightarrow 0$ while the
effective length scale $l_{eff}$ is kept finite. Here $T_{M2}={1\over
4\pi^2l_p^3}$ is the membrane tension and $T_{eff}= {1\over
4\pi^2l_{eff}^3}$ is the finite effective tension of the light
membranes that survive the limit. Note that GM has no coupling
constant and it contains a unique dimensionfull parameter
$l_{eff}$. This is reminiscent of some of the properties of eleven
dimensional M Theory.

The decoupling limit giving rise to GF
is given by
\eqn\GFlim{\eqalign{
g_{\mu\nu}&=\eta_{\mu\nu} \quad \mu,\nu=0,1,\cdots ,5\cr
g_{ij}&={l^6_p\over l^6_{eff}}\delta_{ij} \quad i,j=6,\cdots,10\cr
C_{012345}&=T_{M5}-T_{eff}},}
with  $l_p\rightarrow 0$ while the
effective length scale $l_{eff}$ is kept finite. Here $T_{eff}$
denotes the effective tension of the light five-brane
excitations. Just as GM, GF has no coupling constant and $l_{eff}$ is
the unique dimensionless parameter of the theory.

Similar theories can be obtained from light
Dp-branes, which we will call GDp (Galilean D-p branes). As we will now
see, these theories are connected by dualities which are reminiscent
of the dualities of the fully relativistic theories from which we
obtain these Galilean theories. We first discuss the relation between
supersymmetric Type IIA NRCS and the eleven dimensional theory GM.

\medskip
\noindent
$\underline{\it Strong~ coupling~ dual~ of~ Type~ IIA~ NRCS}$
\medskip

Here we will show that in fact the strong coupling dual to Type IIA
NRCS is described by eleven dimensional GM such that the parameters of
the two theories are related to each other in a similar fashion to the
usual relation in the relativistic setting. 

Consider GM theory \GMlim\ compactified on a circle of proper radius
$R$. The background critical three form potential reduces to a
NS-NS two-form potential
\eqn\pot{
  B_{01}=2\pi R C_{012}.}
Using the usual relation of parameters between M theory and Type IIA
superstring theory
\eqn\pama{
R=g_{s}\sqrt{\ap}\qquad
l_p=g_s^{1/3}\sqrt{\ap},}
we can write $R$ and $l_p$ in the limit \critical\ which defines NRCS,
so that 
\eqn\pamafi{
R=g\sqrt{\alpha_{eff}'}\qquad l_p=g^{1/3}\ap^{1/6}_{eff}\ap^{1/3},}
as $\ap\rightarrow 0$.

Substituting the eleven dimensional limit \GMlim\ in \pot\ we see that
the background NS-NS two-form potential is given by
\eqn\nscri{
2\pi\ap B_{01}=1-{l_p^3\over l_{eff}^3}.}
Using \pamafi\ and comparing with the limit defining NRCS in
\critical , we see that NRCS with coupling constant $g$ and effective string scale
$\alpha_{eff}'$ is equivalent to GM theory on a circle of radius $R$ and
effective Planck scale $l_{eff}$. The parameters are related by
\eqn\pamap{
R=g\sqrt{\alpha_{eff}'}\qquad\qquad l_{eff}=g^{1/3}\sqrt{\alpha_{eff}'}.}
\smallskip

\subsec{GDp and GNSF Theories}

In this subsection we present some generalizations to the construction we made 
for perturbative closed strings in a near critical NS-NS B-field to
Dp-branes in a near critical Ramond-Ramond $p+1$-form background. In
order for the background to affect the Hamiltonian, the spatial
directions of the brane have to be compactified on an orientable $p$-cycle of
space-time. In this case, winding number plays the role of electric
charge in the discussion in \nonr\ and again positivity of the energy
allows only wrapping in one orientation.

The non-relativistic limit that needs to be taken requires
keeping finite the terms in the worldvolume action depending on the
transverse coordinates to the brane and tuning
the background 
field to the tension of the D-brane\foot{This is the analog of the
NRCS limit. There, we scaled $\ap$ and the metric in the transverse
directions to the B field such that terms of the string worldsheet
action depending on the transverse coordinates are kept finite.}. The
limit is given by 
\eqn\limiod{\eqalign{
g_{\mu\nu}&=\eta_{\mu\nu}, \quad \mu,\nu=0,1,\cdots, p\cr
g_{ij}&=\left({\ap\over
\alpha_{eff}'}\right)^2 \delta_{ij}, \quad 
i,j=p+1,\cdots ,9 \cr  
g_s&=\left({\alpha'\over
\alpha_{eff}'}\right)^{{3-p\over 2}}g_p \cr 
C_{01\cdots p}&=T_p-T_p^{eff},}} 
as $\ap\rightarrow 0$.
Here $T_p$  is the tension of the D-brane (which is infinite in the
limit)
\eqn\tens{
T_p={1\over (2\pi)^p g_s \ap^{{p+1\over 2}}}.}
$T_p^{eff}$ is the finite scale of the non-relativistic theory and
$g_p$ the coupling of the theory. The effective tension 
is given by 
\eqn\tensfin{
T_p^{eff}={1\over (2\pi)^p g_p
\alpha_{eff}'^{p+1\over 2}}.} 
In the limit \limiod\  all states of string theory decouple except the
light Dp-branes that have finite proper energy excitations. In
particular, open and closed strings decouple from the low energy
spectrum.

There is a very simple explanation for the limit we take in \limiod . 
Except for the presence of the background gauge field $C_{01\cdots p}$, 
\limiod\ is
the conventional low energy limit that results in a gauge theory on a 
Dp-brane \ref\dkps{M. Douglas, D. Kabat, P. Pouliot, and 
S.H. Shenker, ``D-branes and Short Distances in String Theory,''
Nucl. Phys. B485 (1997) 85, {\tt hep-th/9608024}.}. 
This can be easily recognized by noting that
the Yang-Mills coupling constant is given by $g^2_{YM}\sim g_p\,
{\alpha_{eff}}'^{{p-3\over 2}}$ and that the limit \limiod\
keeps the coupling finite. 
Moreover, the metric for the transverse coordinates $X^i$ 
to the brane is given as in \limiod\ whenever we express them
in terms of the Higgs fields $\Phi^i$ of the gauge
theory as $X^i = \alpha' \Phi^i$ and require that the metric
for $\Phi^i$ remains finite. The low
energy limit is supplemented by turning a near critical background
gauge field which results in light Dp-brane fluctuations.

There is one more theory we can define by tuning a massless gauge
field of string theory, namely the one where the light
excitations are NS five-branes. We will call these theories GNSF
(Galilean Neveu-Schwarz five-brane). Just like NRCS, these theories 
can be obtained as a low energy limit of the different superstring
theories. This low energy limit can also be motivated by considering
the low energy limit on NS five-branes which yields the little string
theories
\lref\seibli{N. Seiberg, ``Matrix Description of M-theory on $T^5$ and
$T^5/Z_2$'', Phys. Lett. B408 (1997) 98, {\tt hep-th/9705221}.}
\seibli .
 The non-relativistic limit is given by
\eqn\limiGNSF{\eqalign{
g_{\mu\nu}&=\eta_{\mu\nu} \quad \mu,\nu=0,1,\cdots, 5\cr
g_{ij}&=\left(g_s \over G\right)^2 \delta_{ij} \quad 
i,j=6,\cdots,9\cr  
B_{012345}&=T_5-T_5^{eff}},} 
as $g_s\rightarrow 0$ while keeping the string scale $\ap$ finite in
the limit. Now the effective tension of the
light NS five-branes is given by
\eqn\effe{
T_5^{eff}={1\over (2\pi)^5 {G}^2{\ap}_{eff}^3}.}
Thus, \limiGNSF\ defines a non-relativistic theory GNSF of light
fluctuations of NS five-branes.

We will now briefly describe the theory one obtains for different
values of $p$. 
\smallskip
\noindent
$\underline{\it Zero~branes}$
\smallskip

For GD0, one may lift the description to
M Theory\foot{For a
lift to eleven dimensions of a similar limit see \gmss .}.
 After a suitable rescaling of energy scales in M Theory, the
M Theory circle goes from a space-like circle to a light-like
circle of finite radius
$R=g_0\sqrt{\alpha_{eff}'}$. Then, if we
consider the  
limit \limiod\ for $N$ D0-branes one obtains a DLCQ description of
M Theory with eleven dimensional Planck length
$l_p=g_0^{1/3}\sqrt{\alpha_{eff}'}$ in a sector
with $N$ units of 
longitudinal momentum.

\smallskip
\noindent
$\underline{\it One~branes}$
\smallskip

For $p=1$, the light states are D1-branes. The strong coupling dual of
the theory of light
D1-branes can be found by using Type IIB S-duality. S-duality maps the
critical R-R background $C_{01}$ to a critical NS-NS background
$B_{01}$. The parameters $\widetilde{g}_s$ and $\widetilde{\alpha}'$ 
of the S-dual theory are given by
\eqn\sdualpara{
\widetilde{g}_s
={1\over g_s}\qquad \widetilde{\ap}=\ap g_s.}
Writing the limit \limiod\ for $p=1$ in terms of the S-dual variables
and comparing with \critical\ 
shows that the S-dual of GD1 is given precisely by NRCS if the
parameters of the two theories are related by
\eqn\paramap{
g={1\over \widetilde{g}}\qquad  \alpha_{eff}'=\widetilde{g}\;
\widetilde{\ap}_{eff}.}
Thus Type IIB NRCS is related to GD1 by a strong-weak coupling duality
which takes the same form as the conventional Type IIB S-duality. A
the end of this section we will realize S-duality of Type IIB NRCS
by studying GM theory compactified on a two-torus \lref\jhs{
J.H. Schwarz, ``The Power of M Theory,'' Phys. Lett. B367 
(1996) 97, {\tt hep-th/9510086}.}\lref\aspin{P.S. Aspinwall, 
``Some Relationships Between Dualities in String Theory,''
Nucl. Phys. Proc. Suppl. 46 (1996) 30,
{\tt hep-th/9508154}.} \refs{\aspin,\jhs}.

\smallskip
\noindent
$\underline{\it Two~branes}$
\smallskip

GD2 can also be lifted to an eleven dimensional picture. The D2-branes
lift to M2-branes and the parameters (the eleventh dimensional proper
radius $R$ 
and the Planck length $l_p$) of M Theory are related to those
of GD2 by
\eqn\paramap{
R={\ap \over \sqrt{\alpha_{eff}'}} g_2 \qquad
l_p=g_2^{1/3}{\alpha'^{2/3}\over \alpha_{eff}'^{1/6}}.}
Moreover, the near critical R-R background lifts to a near critical
background for the three-form of eleven dimensional
supergravity. Therefore, GD2 can be identified with GM on a
transverse circle. Using the parameters in \GMlim\ we see that GM with an
effective Planck scale $l_{eff}$ on a transverse circle of coordinate
size L is  GD2 with coupling $g_2$ and effective string
scale $\alpha_{eff}'$. The parameters are related by
\eqn\mapp{
L=g_2\sqrt{\alpha_{eff}'}\qquad
l_{eff}=g_2^{1/3}\sqrt{\alpha_{eff}'}.}
Thus, the relation between GD2 and GM is reminiscent of the
conventional duality between Type IIA and M Theory.

\smallskip
\noindent
$\underline{\it Three~branes}$
\smallskip

In order to get light D3-branes one must turn on a critical RR
four-form. We can analyze the strong coupling dual of GD3. In fact,
GD3 is self-dual, since S-duality of the underlying string theory maps
the limit \limiod\ to an analogous limit but with a different coupling
constant and effective string scale. Therefore, GD3 with parameters
$g_3$ and $\alpha_{eff}'$ is dual to GD3 with
parameters $\widetilde{g}_3$ and $\widetilde{\ap}_{eff}$. 
The parameters are related by
\eqn\monto{
\widetilde{g}_3={1\over g_3}\qquad
\widetilde{\ap}_{eff}=g_3\; \alpha_{eff}'.}
At the end of this section we realize the S-duality of CD3 from an
eleven dimensional perspective.

\smallskip
\noindent
$\underline{\it Four~branes}$
\smallskip

The GD4 limit can also be lifted to M Theory. The main difference here
is that $g_s\rightarrow \infty$ in the limit. Therefore,
it is required to analyze the configuration in eleven dimensions. The
D4-branes lift to M5-branes wrapped on the M Theory circle. The proper
size of the circle and the corresponding eleven dimensional Planck
length are given by
\eqn\mlimito{
R=g_{4}\sqrt{\alpha_{eff}'}\qquad
l_p=g_{4}^{1/3}{\ap^{1/3}
\alpha_{eff}'^{1/6}}.} 
Since the background RR five-form lifts to a near critical six-form of
eleven dimensional supergravity, one sees that the strong coupling
dual to GD4 is given by GF. Comparing the parameters in \mlimito\ with
those that define GF \GFlim\ one finds that the parameters of the two
theories are related by
\eqn\mlimitofin{ 
R=g_{4}\sqrt{\alpha_{eff}'}\qquad
l_{eff}=g^{1/3}_4 \sqrt{\alpha_{eff}'}.}

\smallskip
\noindent
$\underline{\it Five~branes}$
\smallskip

We now consider the OD5 theory. For $p=5$, one also has to perform
S-duality since in the decoupling 
limit $g_s\rightarrow \infty$. S-duality maps the
D5-branes to NS five-branes. Moreover, the critical RR field gets mapped
to a critical NS-NS electric field for the NS five-branes. Thus we are
led to studying NS five-branes in a critical field. Type IIB
S-duality maps the limit \limiod\ for $p=5$ to a theory 
with string scale and string coupling given by
\eqn\newcoup{
\widetilde{\ap}=g_s\,\ap=g_5\;\alpha_{eff}'\qquad 
\widetilde{g}_s={1\over
g_s}={\ap\over\alpha_{eff}'}{1\over
g_5}.} 
Note that the string scale of the S-dual theory is finite while the
string coupling vanishes. 
This is precisely the limit that defines Type IIB GNSF \limiGNSF . The
parameters of GD5 $(g_5,\ap_{eff})$ are related to those of GNSF
$(G,\widetilde{\ap})$ by 
\eqn\pospa{
\ap_{eff}=G\,\widetilde{\ap}\qquad G={1\over g_5}.}
In the limit that defines GD5, apart from having low energy D5-brane
excitations, there are also finite energy D1-string excitations. These
are identified in the Type IIB GNSF theory with strings of tension
$\widetilde{\ap}^{-1}$ fluctuating inside the five-branes.

\smallskip
\noindent
$\underline{\it Type~ IIA~ Neveu-Schwarz~five-branes}$
\smallskip

Type IIA GNSF has an interesting strong coupling dual. The limit in
\limiGNSF\ can be realized by considering the decoupled theory of light
fluctuating five-branes of M Theory on a transverse circle of proper
size $R$. The   near
critical six-form background of eleven dimensional supergravity
becomes a near critical RR six-form of Type IIA string theory and the
M5-brane becomes a NS five-brane on a transverse circle of proper
radius $R$. The parameters of the two theories are related by
\eqn\mthpar{
R=g_s\sqrt{\ap}\qquad
l_p=g_s^{1/3}\sqrt{\ap}.} 
By comparing the scaling limit \GFlim\ with \limiGNSF\ and using
\mthpar\ one finds the following relation between the effective length
scales of the two theories
\eqn\lebghtsc{
l_{eff}=G^{1/3}\sqrt{\alpha_{eff}'}.}
Moreover, the NS five-branes now sit at a point in the transverse
circle, whose coordinate length is given by
\eqn\cooflo{
L=G\,\sqrt{\ap}.}

We see that the Galilean theories we have found sit on the same
moduli space and satisfy many of the properties of the parent theories
from which we define them by taking a low energy limit. The reduced
number of degrees of freedom that these theories have can be an interesting
avenue in which to study in a simplified setting M Theory dualities.

\medskip
\noindent
$\underline{\it S-duality~ of~ Type~ IIB~ NRCS~ from~ GM}$
\medskip

We first notice that Type IIB NRCS 
can be obtained by considering eleven dimensional GM theory
compactified on a two-torus, where one circle is along the membrane
and the other is transverse to it, and then shrinking the torus.

Let's consider GM theory compactified on a rectangular torus of
coordinate size radii $R_1$ and $R_2$. We will take the circle of
radius $R_1$ to be along the background three-form of supergravity and
the circle of radius $R_2$ to be transverse to it. If we reduce GM on
$R_1$ one gets Type IIA NRCS. The string coupling is given by 
\eqn\coupq{
g_s={R_1\over \sqrt{\ap}}.}
We now perform T-duality along the circle of radius $R_2$. This maps
the limit to a Type IIB set-up. T-duality inverts the proper size of
the circle one T-duals along and changes the dilaton in the
conventional fashion. The new string coupling is given by
\eqn\newcopaf{
g_s'={R_1\over R_2}\sqrt{{\alpha_{eff}'\over \ap}}.}
Therefore, comparing with \critical\ we see that this compactification
of GM leads to Type IIB NRCS with coupling constant
\eqn\clod{
g={R_1\over R_2}}
compactified on a transverse circle of coordinate size
${{\alpha_{eff}'\over R_2}}$. Therefore, one gets Type IIB NRCS from GM in
the limit that the coordinate area of the torus vanishes at fixed
ratio ${R_1\over R_2}$.

However, one could have chosen to reduce GM on the circle of
coordinate radius $R_2$. As we showed, this leads to CD2 theory on a
parallel circle of coordinate size $R_1$. The Type IIA coupling is
given by
\eqn\iiapo{
\widetilde{g}_s={\sqrt{\widetilde{\ap}}\over \widetilde{\ap}_{eff}}R_2.}
One can perform a T-duality
transformation along the circle of radius $R_1$ such that we get a low
energy limit 
in the Type IIB superstring where the string coupling is given by
\eqn\newlool{
\widetilde{g}_s'={{\widetilde{\ap}}\over
\widetilde{\ap}_{eff}}{R_2\over R_1}}
and the circle is of coordinate size ${\widetilde{\ap}_{eff}\over
R_1}$. Thus, by looking at \limiod\ for $p=1$ we see that one gets CD1
theory with coupling 
\eqn\copligj{
\widetilde{g}_{1}={{R_2\over R_1}}.}

Therefore, S-duality of Type IIB NRCS can be understood from an eleven
dimensional perspective as the symmetry that exchanges the two circles
when one considers GM theory compactified on a two-torus \refs{\aspin,\jhs}.

\medskip
\noindent
$\underline{\it S-duality~ of~ GD3~ from~ GF}$
\medskip

Consider a single M5 brane compactified on a rectangular two-torus of
radii $R_1$ and $R_2$ along the background directions in the
decoupling limit \GFlim . If we treat the 
circle of radius $R_1$ to be the one that results in GD4, then the
parameters of GD4 are given by
\eqn\mlimitofina{
R_1={g}_{4}\sqrt{{\ap}_{eff}}\qquad
l_{eff}={g}^{1/3}_4 \sqrt{{\ap}_{eff}}.}

One may perform T-duality along the circle of radius $R_2$. Then, we
obtain the Galilean theory of light D3-branes. The string coupling
after T-duality is given by
\eqn\newsetr{
{g}_s'={g}_s{\sqrt{\ap}\over
R_2}={g}_4 {\sqrt{{\ap}_{eff}\over R_2}}}
Comparing \newsetr\ with \limiod\ for $p=3$ and using \mlimitofina\
one finds that the effective coupling of the theory of light D3-branes
is 
\eqn\copnew{
{g}_3={R_1\over R_2},}
which is reminiscent of the relation between M Theory on a two-torus
and Type IIB 
string theory. 

If one reduces first on the circle of radius of $R_2$ and then
performs T-duality along the circle of radius $R_1$ one again obtains
GD3 but with the inverse coupling. Therefore, S-duality of GD3 follows
from the interchange of the two circles of the two-torus in the GF
realization of GD3.

\bigskip
\bigskip
\bigskip

\noindent
\centerline{\bf Acknowledgments}

\bigskip
We would like to thank O. Bergman, J. Maldacena, T. Mehen, J. Schwarz,
M. Wise, and E. Witten for useful comments. 
This research is supported in part by
the DOE grant DE-FG03-92ER40701 and
the Caltech Discovery Fund. 

\vfill\eject

\listrefs
\end